\documentclass[12pt]{article}
\usepackage{a4wide}
\usepackage{cite}
\usepackage{pslatex}
\usepackage{graphicx}
\usepackage[latin1]{inputenc}
\usepackage[T1]{fontenc}

\usepackage{color}
\usepackage{colordvi}

\newcommand\fullversion[1]{}

\def\entry#1#2{\parbox[t]{5.5cm}{\it %
    #1:}\hspace*{0.35cm}\parbox[t]{11.0cm}{#2}\\[0.1cm]}

\def\msbar{\ensuremath{\overline{\mbox{MS}}}}

\def\shat{{\hat s}}
\def\mt{{m_t}}
\def\mts{{m_t^2}}
\def\muf{{\mu_f}}
\def\mur{{\mu_r}}
\def\mufs{{\mu_f^2}}
\def\murs{{\mu_r^2}}
\def\mmu{{\overline{m}(\mur)}}
\def\mm{{\overline{m}(\overline{m})}}

\def\as{{\alpha_s}}
\def\nf{{n_f}}
\def\b#1{\beta_{#1}}

\def\fqqn{f_{q\bar{q}}^{(0)}}
\def\fgqn{f_{gq}^{(0)}}
\def\fggn{f_{gg}^{(0)}}

\def\fqqon{f_{q\bar{q}}^{(10)}}
\def\fqqoo{f_{q\bar{q}}^{(11)}}
\def\fqqtn{f_{q\bar{q}}^{(20)}}
\def\fqqto{f_{q\bar{q}}^{(21)}}
\def\fqqtt{f_{q\bar{q}}^{(22)}}

\def\fgqon{f_{gq}^{(10)}}
\def\fgqoo{f_{gq}^{(11)}}
\def\fgqtn{f_{gq}^{(20)}}
\def\fgqto{f_{gq}^{(21)}}
\def\fgqtt{f_{gq}^{(22)}}

\def\fggon{f_{gg}^{(10)}}
\def\fggoo{f_{gg}^{(11)}}
\def\fggtn{f_{gg}^{(20)}}
\def\fggto{f_{gg}^{(21)}}
\def\fggtt{f_{gg}^{(22)}}

\def\fij#1{{f_{ij}^{(#1)}}}
\def\fijon{f_{ij}^{(10)}}
\def\fijoo{f_{ij}^{(11)}}
\def\fijtn{f_{ij}^{(20)}}
\def\fijto{f_{ij}^{(21)}}
\def\fijtt{f_{ij}^{(22)}}

\def\z#1{\zeta(#1)}
\def\d#1{{d_{#1}}}

\def\Lij{{{\cal L}_{ij}}}
\def\sigmaij#1{{\hat\sigma_{ij}^{(#1)}}}
\def\sig#1{{\sigma^{(#1)}}}
\def\Dsigma#1{\left.{d\sigma^{(#1)}(\mt)\over d\mt}\right|_{\mt=\mm}}
\def\DDsigma#1{\left.{d^2\sigma^{(#1)}(\mt)\over d\mt^2}\right|_{\mt=\mm}}

\newcommand{\Liz}{{\rm Li}_2}
\newcommand{\ha}{\ln\left(\frac{1+\beta}{1-\beta}\right)}
\newcommand{\hb}{\ln \left(\frac{\rho}{4\*\beta^2}\right)}
\newcommand{\argp}{\frac{1+\beta}{2}}
\newcommand{\argm}{\frac{1-\beta}{2}}
\newcommand{\argpp}{\frac{2\*\beta}{1+\beta}}
\newcommand{\argmm}{\frac{-2\*\beta}{1-\beta}}

\def\nn{\nonumber}

\begin{document}

\begin{titlepage}
\noindent
DESY 10-091 \\
HU-EP-10-33 \\
SFB/CPP-10-60
\vspace{1.3cm}

\begin{center}
  {\bf 
\Large
    -- HATHOR  --\\[2ex]
\large
{\Large{HA}}dronic {\Large{T}}op and {\Large{H}}eavy quarks cr{\Large{O}}ss section calculato{\Large{R}}
    }
  \vspace{1.5cm}

  {\large
  M. Aliev$^{\,a}$, H. Lacker$^{\,a}$, U. Langenfeld$^{\,a}$, 
  S. Moch$^{\, b}$, \\[2ex] 
  P. Uwer$^{\,a}$  and M. Wiedermann$^{\,a}$}\\
  \vspace{1.2cm}

  {\it 
    $^a$Humboldt-Universit\"at zu Berlin, Institut f\"ur Physik\\
    Newtonstra{\ss}e 15, D-12489 Berlin, Germany
    \vspace{0.2cm}
    
    $^b$Deutsches Elektronensynchrotron DESY \\
    Platanenallee 6, D--15735 Zeuthen, Germany}
  \vspace{1.4cm}

  {\large\bf Abstract}
  \vspace{0.5cm}

  \parbox{14.5cm}{%
    We present a program to calculate the total cross section for top-quark
    pair production in hadronic collisions. 
    The program takes into account recent theoretical developments  
    such as approximate next-to-next-to-leading order perturbative QCD corrections 
    and it allows for studies of the theoretical uncertainty by 
    separate variations of the factorization and renormalization scales. 
    In addition it offers the possibility to obtain the cross section as a function of the running top-quark mass. 
    The program can also be applied to a hypothetical fourth quark family provided the QCD couplings are standard. 
    }
\end{center}
\end{titlepage}

%
%
\section*{Program summary}
\entry{Title of program}{Hathor}

\entry{\it Version}{1.0}

\entry{Catalogue number}{}

\entry{Program summary URL}{\tt
  http://www.physik.hu-berlin.de/pep/tools \\
  http://www-zeuthen.desy.de/\~{}moch/hathor
  }

\entry{E-mail}{\tt sven-olaf.moch@desy.de,
  peter.uwer@physik.hu-berlin.de}

\entry{License}{GNU Public License}

\entry{Computers}{Standard PCs (x86, x86\_64 processors)}

\entry{Operating system}{
  Linux}

\entry{Program language}{
  C++, fortran, Java}

\entry{Memory required to execute}{
  256 MB}

\entry{Other programs called}{
  None.}

\entry{External files needed}{
  Interface to LHAPDF for the user's choice of parton distribution functions, 
  see {\tt http://projects.hepforge.org/lhapdf/}.}

\entry{Keywords}{
  Top-quarks, total cross section, QCD, radiative corrections, running mass.}

\entry{Nature of the physical problem}{
  Computation of total cross section in perturbative QCD.}

\entry{Method of solution}{
  Numerical integration of hard parton cross section convoluted with parton
  distribution functions.}

\entry{Restrictions on complexity of the problem}{
  None}

\entry{Typical running time}{
  A few seconds to a few minutes on standard desktop PCs or notebooks,
  depending on the chosen options.}

\newpage
%
%
\section{Introduction}
\label{sec:intro}
The top-quark is the heaviest elementary particle in nature discovered so far. 
As a consequence of its large mass close to the scale of the electroweak symmetry breaking 
it has remarkable properties making it a distinct research object. 
For example the short lifetime does not allow the formation of hadronic bound-states. 
Rather, the top-quark decays before it hadronizes, a fact often referred to colloquially 
as the top-quark behaving like a quasi-free quark. 
Non-perturbative effects are thus essentially cut-off by the short lifetime and, 
as an important consequence, the polarization of top-quarks can be studied 
through the parity violating decay into a $W$-boson and a bottom quark. 
The top-quark also provides an interesting environment for precision tests of
the Standard Model (SM) and possible extensions, e.g., by constraining the
allowed range for the Higgs mass.

A mandatory ingredient for top-quark physics at hadron colliders are precise theoretical predictions to compare with. 
Current Tevatron measurements and the perspectives at LHC, 
i.e., a measurement of the top-quark pair cross section with an uncertainty of
the order of 5\% only set the target for theoretical predictions of the production process. 
Clearly such an accuracy needs to include quantum corrections. 

Within Quantum Chromodynamics (QCD) radiative corrections were calculated some time ago 
to next-to-leading order (NLO) first considering unpolarized 
top-quark production~\cite{Nason:1987xz,Beenakker:1988bq} 
and later including spin information~\cite{Bernreuther:2004jv}. 
In the former case, also completely analytical results have recently been provided~\cite{Czakon:2008ii}.
Beyond the NLO accuracy in QCD various sources of possible improvements have been identified. 
Large logarithmic corrections due to soft gluon emission were investigated and
resummed at the next-to-leading-logarithmic (NLL) accuracy~\cite{Kidonakis:1997gm,Bonciani:1998vc} 
and recently improved to include also 
the next-to-next-to-leading-logarithmic (NNLL) corrections~\cite{Moch:2008qy,Czakon:2009zw,Beneke:2009rj}.
Alternatively, resummation has also provided the means to construct parts of the full 
next-to-next-to-leading (NNLO) fixed order results~\cite{Moch:2008qy,Moch:2008ai,Beneke:2009ye}, 
which can be supplemented by including all Coulomb type corrections~\cite{Beneke:2009ye}  
and also the full scale dependence at NNLO accuracy~\cite{Moch:2008qy,Langenfeld:2009wd}.
With a target precision for the total cross section at the few per cent level also bound state effects
from the resummation of Coulomb type corrections~\cite{Hagiwara:2008df,Kiyo:2008bv} 
as well as electro-weak radiative corrections 
at NLO~\cite{Beenakker:1993yr,Bernreuther:2006vg,Kuhn:2006vh} need to be considered.

The compilation of all these results is in principle straight forward given the extensive literature on the subject. 
However, no publicly available program exists so far which contains the latest theoretical developments.
The ``modus operandi'' of the past was that predictions were updated by theorists from time to time 
taking into account new theoretical improvements and/or new sets of parton distribution functions (PDFs). 
The aim of the present paper is to provide a program for the computation of the 
top-quark pair cross section including state of the art theory.
As such it can serve as a reference for future cross section calculations. 
The program Hathor includes perturbative QCD corrections at higher orders in the different approximations 
along with options allowing also a detailed study of the theoretical uncertainties. 
Moreover, it provides the possibility to compute the total cross section not
only in the commonly adapted pole mass scheme but also in terms of the \msbar\ mass 
a choice recently employed for top-quark pair production in hadronic collisions 
for the first time~\cite{Langenfeld:2009wd,Moch:2010rh}.
Finally, the aim of this publication is not only to provide a tool for cross section calculations 
but also to facilitate experimental analyses. 
To that end the package contains in addition to the stand alone program 
also a small library that can be easily integrated into existing code.

The outline of the article is a follows. 
In the next Section we briefly discuss the theoretical foundations 
as well as the procedure to convert to the \msbar\ mass.
Sections~\ref{sec:installation} and~\ref{sec:description} 
contain installation details and the program description while usage and examples are given in Section~\ref{sec:usage}.
We end with conclusions in Section~\ref{sec:conclusions}.
All formulae as implemented in Hathor are collected in the Appendices~\ref{sec:msbar} and~\ref{sec:fits}.

%
%
\section{Methods}
\label{sec:methods}
The hadronic cross section for top-quark pair production is obtained
from the convolution of the factorized partonic cross section 
$\hat \sigma_{ij}$ with the parton luminosities $\Lij$:
\begin{equation}
  \label{eq:totalcrs}
  \sigma_{h_1 h_2 \to {t\bar t X}}(S,\mt) = 
  \sum_{i,j}
  \int\limits_{4\mts}^{S}\,
  ds \,\, \Lij(s, S, \muf)\,\,
  \hat \sigma_{ij}(s,\mt,\as(\mur),\muf) 
  \, , 
\end{equation}
\begin{equation}
  \label{eq:partonlumi}
  \Lij(s,S,\muf) = 
  {1\over S} \int\limits_s^S
  {d\shat\over \shat} 
  f_{i/h_1}\left({\shat \over S},\muf\right) 
  f_{j/h_2}\left({s \over \shat},\muf\right).
\end{equation}
Here $S$ denotes the hadronic center-of-mass energy squared, $\mur$,
($\muf$) denotes the renormalization (factorization) scale 
and the functions $f_{i/h_{1,2}}(x,\muf)$ are the PDFs describing the probability to find a parton 
of type $i$ with a momentum fraction between $x$ and $x+dx$ in the hadron $h_k$. 
The QCD coupling constant $\as(\mur)$ is evaluated at the scale $\mur$. 
In the following we use $\as$ in the scheme with $\nf$ light flavors. 
For top-quark production, the running is thus determined by the five light
flavors $u,d,c,s,b$ which we treat as massless. 
The top-quark mass $\mt$ appearing in Eq.~(\ref{eq:totalcrs}) 
is the mass renormalized in the on-shell (pole-mass) scheme. 

In perturbative QCD the partonic cross section 
$\hat \sigma_{ij}(s,\mt,\as,\muf)$ is expanded in the QCD coupling constant 
up to NNLO:
\begin{equation}
  \hat \sigma_{ij} \,=\, a_s^2\,  \sigmaij0(s,\mt) 
  + a_s^3\,  \sigmaij1(s,\mt,\mur,\muf) 
  + a_s^4\,  \sigmaij2(s,\mt,\mur,\muf) 
  \,+\, O(a_s^5)
  \, ,
\end{equation}
with $a_s = \as/\pi$.

In leading-order (LO) only the parton channels $q \bar q$ and $gg$ contribute 
and the respective Born cross sections are given by:
\begin{eqnarray}
  \label{eq:LOqq}
   \hat\sigma_{qq}^{(0)} &=& {4\pi^3\over 27}  {1\over s}
  \beta\*(3-\beta^2),\\  
  \label{eq:LOgg}
  \hat \sigma_{gg}^{(0)} &=&
  {\pi^3\over 48}\*{1\over s}\*\left\{
    (33-18\*\beta^2+\beta^4)\*\ln\left({1+\beta\over 1-\beta}\right)
    -59\*\beta+31\*\beta^3 \right\},
\end{eqnarray}
with $\beta = \sqrt{1-\rho}$ and $\rho = 4\mts / s$.
Starting from NLO also the $gq$ and $g\bar q$ channels contribute. 
In Ref.~\cite{Nason:1987xz} (and in many subsequent publications) an alternative 
decomposition was used in terms of so-called scaling functions $f_{ij}$:
\begin{equation}
  \label{eq:DefinitionScalingFunctions}
   \hat \sigma_{ij} \,=\, 
   {\as^2\over \mts} \left\{
   \fij0(\rho) 
   + 
   4\pi \as \fij1(\rho,\muf/\mt,\mur/\muf)
   +
   (4\pi \as)^2 \fij2(\rho,\muf/\mt,\mur/\muf) 
   \,+\, O(\as^3)
   \right\}
     \, .
\end{equation}
Since the scaling functions are dimensionless they depend only on $\rho$ 
and the ratios $\muf/\mt$ and $\mur/\muf$.
The full renormalization and factorization scheme dependence can be constructed 
using the renormalization group equation, the standard evolution equations of the PDFs 
and information about lower orders, i.e., up to NNLO knowledge of $\fij0(\rho)$
and $\fij1(\rho,1,1)$ is sufficient.
The general structure can be written in the following form 
\begin{eqnarray}
\label{eq:scale-dep-1}
  \fij1(\rho,\muf/\mt,\mur/\mt) &=& 
  \fijon +  L_M \fijoo + 2 \b0 L_R \fij0
  \, ,
  \\
\label{eq:scale-dep-2}
  \fij2(\rho,\muf/\mt,\mur/\mt) &=& 
  \fijtn + L_M \fijto + L_M^2 \fijtt + 3\b0 L_R \fijon + 3 \b0 L_R L_M \fijoo 
  \nonumber\\
  & & + 2 \b1 L_R \fij0 + 3 \b0^2 L_R^2 \fij0 
  \, ,
\end{eqnarray}
with $ij = \{q{\bar q}, gg\}$ and we abbreviate $L_M = \ln(\mufs / \mts)$ and $L_R = \ln(\murs / \mufs)$.
The scale dependence in the $gq$ ($g \bar q$) channel can be easily derived from the above 
realizing that $\fgqn = 0$ 
so some terms in Eqs.~(\ref{eq:scale-dep-1}) and (\ref{eq:scale-dep-2}) are simplify absent.
In the conventions used here the coefficients of the beta-function $\b0, \b1$ are given by
\begin{equation}
  \label{eq:defbeta}
  \b0 = {1\over (4\pi)^2 } \left(11-{ 2\over 3} \nf\right)\, ,
  \qquad\qquad 
  \b1 = {1\over (4\pi)^4 } \left(102-{38\over 3} \nf\right) 
  \, .
\end{equation}

The Born contributions have been presented in Eqs.~(\ref{eq:LOqq}),~(\ref{eq:LOgg}) 
and at present also the complete NLO corrections are known, 
i.e., the functions $\fijon$ and  $\fijoo$ in Eq.~(\ref{eq:scale-dep-1}).
A complete NNLO calculation for the total cross section is not yet available, 
since $\fij2(\rho,\muf/\mt,\mur/\mt)$ in Eq.~(\ref{eq:scale-dep-2}) is missing the contribution $\fijtn$ 
while $\fijto$ and $\fijtt$ have been obtained from renormalization group arguments as mentioned above.
However, exact expressions for $\fijtn$ in the limit $\rho \to 1$ based on soft-gluon resummation have been derived 
and provide the foundation for approximate NNLO results of $\sigma_{h_1 h_2 \to {t\bar t X}}$.

The central physics questions to be addressed can be phrased as follows: 
\begin{itemize}
\item 
How large is the total cross section $\sigma_{h_1 h_2 \to {t\bar t X}}$ at a given order in perturbation theory ?

\item
Given a computation of the total cross section according to Eq.~(\ref{eq:totalcrs}) 
what is the associated theoretical uncertainty ?
\end{itemize}
In order to address these issues the package Hathor has different production models implemented which 
are accessible to the user as options. 
In the following we briefly describe these options as far as the underlying physics is concerned. 
To be self-consistent and for easier reference all necessary theory input, 
e.g., the scaling functions has been collected in Appendix~\ref{sec:fits}.
For details of how to access these options when running Hathor we refer to the
next Sections~\ref{sec:description} and \ref{sec:usage}.

\subsubsection*{Option {\tt LO}}
The option {\tt LO} provides a rough estimate although with large theoretical uncertainties
which will receive sizable corrections at higher orders. 
This option uses the Born cross sections 
of Eqs.~(\ref{eq:LOqq}),~(\ref{eq:LOgg}) (see also Eqs.~(\ref{eq:fqq0})--(\ref{eq:fgg0})).

\bigskip
\subsubsection*{Option {\tt NLO}}
The option {\tt NLO} is the first instance where a meaningful theoretical uncertainty can be quoted 
in perturbation theory.
This option employs the complete NLO QCD corrections~\cite{Nason:1987xz,Beenakker:1988bq}. 
All scaling functions $\fijon$ are given as accurate
fits~\cite{Langenfeld:2009wd} based on the recently published analytic results~\cite{Czakon:2008ii},
(see also Eqs.~(\ref{eq:fqq10})--(\ref{eq:fgg10})).

\bigskip
\subsubsection*{Option {\tt NNLO}}
The option {\tt NNLO} is required whenever predictions with an uncertainty of better than $O(10) \%$ are needed.
This option is based on the known threshold enhancement due to soft gluon emission, 
i.e., complete tower of Sudakov logarithms at NNLO accuracy, supplemented by all Coulomb type corrections~\cite{Beneke:2009ye}  
and also the full scale dependence~\cite{Moch:2008qy,Langenfeld:2009wd} (see Eqs.~(\ref{eq:f21qq})--(\ref{eq:f22gg})). 
This ansatz provides a good approximation for the total cross section~\cite{Moch:2008qy,Moch:2008ai}, 
a fact which is supported by the observation that the QCD corrections to top-quark pair production in association with an
additional jet are small~\cite{Dittmaier:2007wz,Dittmaier:2008uj,Melnikov:2009dn}.

Thus, using the results of Refs.~\cite{Langenfeld:2009wd,Beneke:2009ye}, we have for the functions $\fijtn$: 
\begin{eqnarray}
\label{eq:fqq20}
\fqqtn &=& {\fqqn \over (16 \pi^2)^2} \*
  \biggl\{
         {8192 \over 9} \* \ln^4 \beta
       + ( - 1505.1589 + 37.925926 \* \nf ) \* \ln^3 \beta
\\ &&
       + \biggl( 1046.4831 - 90.838135 \* \nf - 140.36771 \* {1 \over \beta} \biggr) \* \ln^2 \beta
\nonumber\\ &&
       + \biggl( 
         249.67547 + 55.776275 \* \nf 
       + ( 54.038454 - 4.3864908 \* \nf ) \* {1 \over \beta} \biggr) \* \ln \beta
\nonumber\\ &&
       + 3.6077441 \* {1 \over \beta^2} 
       + ( - 5.2728242 + 1.8447758 \* \nf ) \* {1 \over \beta} 
       + C^{(2)}_{q{\bar q}}
    \biggr\}
\, ,
\nonumber\\
\label{eq:fgq20}
\fgqtn &=& {\beta^3 \over (16 \pi^2)^2} \* {65 \*\pi \over 54} \* \ln^3 (8 \* \beta^2)
\, ,
\\
\label{eq:fgg20}
\fggtn &=& {\fggn \over (16 \pi^2)^2} \*
  \biggl\{
         4608 \* \ln^4 \beta
       + ( - 2321.5810 + 85.333333 \* \nf ) \* \ln^3 \beta
\\ &&
       + \biggl( - 315.57218 - 119.35529 \* \nf + 496.30011 \* {1 \over \beta} \biggr) \* \ln^2 \beta
\nonumber \\ &&
       + \biggl( 
         2346.8889 + 21.969529 \* \nf 
       + ( 286.67132 + 6.8930570 \* \nf ) \* {1 \over \beta} \biggr) \* \ln \beta
\nonumber \\ &&
       + 68.547138 \* {1 \over \beta^2} 
       + ( - 3.7910584 - 0.96631115 \* \nf ) \* {1 \over \beta} 
       + C^{(2)}_{gg}
    \biggr\}
\, ,
\nonumber 
\end{eqnarray}
where the unknown functions $C^{(2)}_{q{\bar q}}$ and $C^{(2)}_{gg}$ 
in Eqs.~(\ref{eq:fqq20}) and (\ref{eq:fgg20}) parametrize the contributions
which are not enhanced in the threshold region, i.e., $O(\beta^0)$. 
The $gq$-channel, that is $\fgqtn$ in Eq.~(\ref{eq:fgq20}), 
is additionally suppressed near threshold with corrections of order $O(\beta^3 \ln^2(\beta) )$. 

In summary, the option {\tt NNLO} (which has been used e.g., 
for the phenomenological studies of Ref.~\cite{Langenfeld:2009wd}) uses all presently available information at NNLO.
In this way, it attempts to construct the relevant parts of the complete NNLO corrections. 
Necessarily, the small associated theoretical uncertainty~\cite{Moch:2008qy,Moch:2008ai} 
due to scale variation ($\mur$ and $\muf$) estimates effects beyond NNLO.
An additional systematical uncertainty on the quality of the approximate NNLO result can be 
quantified by varying the constants $C^{(2)}_{q{\bar q}}$ and $C^{(2)}_{gg}$ 
in a reasonable range comparable to the size of the other coefficients 
in Eqs.~(\ref{eq:fqq20}) and (\ref{eq:fgg20}), i.e., $C^{(2)}_{ij}= \pm O(100)$. 
The default value is $C^{(2)}_{ij}=0$.

\bigskip
\subsubsection*{Option {\tt LOG\_ONLY}}
The option {\tt LOG\_ONLY} is also motivated by the idea of soft gluon enhancement near threshold.
It emerged as a conservative definition of the theoretical uncertainty in a comparison of different 
approaches to incorporate dominant terms beyond NLO to a certain logarithmic accuracy.
In particular threshold resummation at NLL accuracy, performed as in Refs.~\cite{Bonciani:1998vc,Cacciari:2008zb} 
which typically proceeds in Mellin-space, see Eq.~(\ref{eq:mellindef}), 
has been tested against an expansion in powers of $\ln^k(\beta)$ in momentum
space as advocated in Ref.~\cite{Moch:2008qy,Moch:2008ai}.
This comparison has yielded satisfactory agreement, because resummation beyond NLL, 
i.e., at NNLL accuracy has only a minor effect~\cite{Moch:2008qy}.

The option {\tt LOG\_ONLY} as discussed below is based on work with CCMMMNU~\cite{ccmmmnu:2010}. 
It is a genuine NLO approach with logarithmic improvement near threshold and 
scale variations in $\mur$ and $\muf$ (with a constraint on the ratio of $\mur/\muf$) estimate effects beyond NLO. 
Being a conservative approach the resulting theoretical uncertainty is
necessarily larger than in the option {\tt NNLO}.

Let us briefly mention the essential technical points. 
In Refs.~\cite{Bonciani:1998vc,Cacciari:2008zb} 
the logarithmic enhancement is constructed from the $\ln(N)$ terms in Mellin space where the resummation is usually performed. 
The transformation from momentum ($\rho-$) space to Mellin ($N$-) space is given by
\begin{equation}
\label{eq:mellindef}
  \sigma(N) = \int_0^1\, d\rho\, \rho^{N-1} \sigma(\rho) 
  \, ,
\end{equation}
where $\rho = 4 \mts/s$.
The important feature of Eq.~(\ref{eq:mellindef}) to realize is that beyond logarithmic accuracy 
the momentum space and the Mellin-space expressions do differ by terms 
which are not enhanced in $\beta$ or, respectively power-suppressed in $1/N$.
Any difference could be included in a choice of $C^{(2)}_{ij}$ in Eqs.~(\ref{eq:fqq20}), (\ref{eq:fgg20}). 

Option {\tt LOG\_ONLY} has to be used with the Option {\tt NNLO}. 
It applies Eqs.~(\ref{eq:fqq20}) and (\ref{eq:fgg20}),
but truncates the function $\fij0$ beyond NLO to is leading term in $\beta$ (cf. Eqs.~(\ref{eq:LOqq}),~(\ref{eq:LOgg})) and, 
for consistency neglects the $gq$-channel beyond NLO ($\fgqtn$ and Eqs.~(\ref{eq:f21gq}) and (\ref{eq:f22gq})).
Also for the scale dependent part, the option {\tt LOG\_ONLY} keeps only
terms which are logarithmically enhanced in the threshold region.
In this case the functions $\fijto$ and $\fijtt$ in Eqs.~(\ref{eq:f21qq})--(\ref{eq:f22gg})) 
are truncated to logarithmic accuracy.
Again, one could also vary the constants $C^{(2)}_{q{\bar q}}$ and $C^{(2)}_{gg}$ 
in a range $C^{(2)}_{ij}= \pm O(100)$ to test for additional systematical uncertainties.
The default value is $C^{(2)}_{ij}=0$. 

\bigskip
\subsubsection*{Option {\tt MS\_MASS}}
The option {\tt MS\_MASS} allows for the computation 
of the total cross sections as a functions of the running mass in the \msbar\ scheme.
In a nut-shell this is based on the replacement $\mt \to \mmu$ (see Eq.~(\ref{eq:pole-to-msbar}))
in the expression for $\sigma_{h_1 h_2 \to {t\bar t X}}$ in Eq.~(\ref{eq:totalcrs}). 
The option {\tt MS\_MASS} can be applied together with options {\tt LO}, {\tt NLO} and {\tt NNLO}.

Let us briefly discuss the main motivation for this option.
So far the mass used in all formulae is given as the so-called on-shell or pole-mass 
which is defined as the location of pole of the quark propagator calculated order-by-order in perturbation theory. 
It is well known that the pole-mass is not a well defined concept in QCD
since quarks do not appear as asymptotic states in the quantum field
theoretical description of the strong interaction owing to confinement. 
In other words, the quark propagator does not have a pole in full QCD. 
A more quantitative analysis leads to the conclusion that the pole-mass 
is intrinsically uncertain of the order of $\Lambda_{QCD}$ (see e.g.~\cite{Bigi:1994em,Smith:1996xz}).
Since in perturbation theory the pole-mass can be expressed in terms
of a short distance mass like for example the running mass which is
free from non-perturbative effects it is advantageous to translate the cross
section predictions from the on-shell scheme to the \msbar\  mass scheme. 
As a benefit, the convergence of the perturbative series is significantly improved 
when the running mass is used and the extracted numerical value of the
top-quark mass is very stable under higher order corrections. 
These observations were first pointed out in Ref.~\cite{Langenfeld:2009wd}.

In the Hathor program the conversion 
$\sigma_{h_1 h_2 \to {t\bar t X}}(S,\mt) \to \sigma_{h_1 h_2 \to {t\bar t X}}(S,\mmu)$ 
has been realized now in an easy way allowing a direct evaluation of the cross section 
using the running mass (see also~\cite{Moch:2010rh}).
All details are deferred to Appendix~\ref{sec:msbar}.

\bigskip
\subsubsection*{Option {\tt PDF\_SCAN}}
The option {\tt PDF\_SCAN} allows for the automated computation of PDF uncertainties.
In the default setting of the Hathor package the PDFs can be accessed with the LHAPDF library~\cite{Whalley:2005nh,lhapdf}. 
A prerequisite for this option is, of course, that the respective PDF provides a set of error PDFs.
There are, however, different conventions with respect to PDF uncertainties.

For instance, there exists the convention of asymmetric uncertainties, a choice 
adopted by e.g. MSTW~\cite{Martin:2009iq} and CTEQ~\cite{Nadolsky:2008zw}.
Here the error PDFs come in $n_{PDF}$ pairs $(\sigma_{k,+},\sigma_{k,-})$, 
where the first element of the pair describes the error of the corresponding parameter 
in the '$+$'-direction, the second the one in the '$-$'-direction. 
Then, for a given PDF set with a central fit resulting in a cross section $\sigma_0$
the systematic uncertainty $\Delta \sigma_\pm$ is estimated by (see e.g.~\cite{Campbell:2006wx}),
\begin{equation}
  \label{eq:asympdferr}
  \Delta \sigma_\pm = 
  \frac{1}{2} \, \sqrt {\sum_{k=1,n_{PDF}} \, 
    ({\rm max}(0,\pm \sigma_{k,+} \mp \sigma_0, \pm \sigma_{k,-} \mp \sigma_0))^2}
  \, .
\end{equation}
Eq.~(\ref{eq:asympdferr}) is the default of the Hathor package when using the option {\tt PDF\_SCAN}. 
Following standard conventions the PDF uncertainty should be linearly added to the theoretical uncertainty 
from scale variations (parameterizing uncalculated higher orders).

Other PDF sets, e.g. ABKM~\cite{Alekhin:2009ni}, employ the convention of symmetric uncertainties, 
where the $n_{PDF}$ elements each describe the (symmetric) '$\pm$'-variation.
In this case, the quadratic uncertainty $\Delta \sigma_\pm$ is obtained by adding the individual errors quadratically
in the standard manner, 
\begin{equation}
  \label{eq:sympdferr}
  \Delta \sigma_\pm = 
  \sqrt {\sum_{k=1,n_{PDF}} \, (\sigma_{k} - \sigma_0)^2}
  \, ,
\end{equation}
and the option \verb+PDF_SCAN+ has to be combined with the additional option \verb+PDF_SYM_ERR+. 

Finally, there exist PDF sets, e.g.~\cite{Alekhin:2000ch,Ball:2010de} 
which simply return a number $n_{PDF}$ of best fits, where typically $n_{PDF} \simeq O(100)$.
Then, the PDF uncertainty of the cross section $\sigma$ is estimated 
by computing it with each of the best fits and taking the standard statistical average.
In such cases, the option {\tt PDF\_SCAN} cannot be used for an automated computation of the PDF uncertainty.
Within Hathor, it of course, always possible for the user to provide own code for the evaluation of the PDF uncertainy.

If, however, a PDF set provides different values of the strong coupling $\alpha_s$ for
different error PDFs, this is automatically taken into account.

\bigskip
\bigskip

Before continuing with the description of the Hathor program, let us
mention that the package offers the possibility for several extensions in the future.
With an experimental precision of 5\% as envisaged at LHC 
the electro-weak radiative corrections at 
NLO~\cite{Beenakker:1993yr,Bernreuther:2006vg,Kuhn:2006vh} have to be taken into account.
While for the Tevatron they turn out to be small (less than $1 \%$) 
they are of the order of $2 \%$ at LHC with a slight dependence on the Higgs mass.
Electro-weak NLO corrections can be included in a similar manner 
as the higher order QCD corrections, i.e., with the help of accurate fits.

Also the treatment of QCD radiative corrections leaves room for improvement,
e.g. by incorporating bound state effects from the resummation of 
Coulomb type corrections~\cite{Hagiwara:2008df,Kiyo:2008bv}.
Finally, the design of the Hathor package can in principle also accommodate related approaches 
for the computation of the total top-quark pair cross section beyond NLO, 
for instance those based on soft gluon enhancement in differential kinematics~\cite{Kidonakis:2008mu,Ahrens:2010zv} 
(see Ref.~\cite{Kidonakis:2001nj} for earlier work).
Such extrapolations of large logarithms from soft gluon emission 
in a differential variable (e.g. the top-quark pair-invariant mass) 
to the full partonic phase space introduce systematic uncertainties 
and require a detailed comparison to an inclusive approach such 
as in Eqs.~(\ref{eq:fqq20})--(\ref{eq:fgg20}).

%
%
\section{Installation}
\label{sec:installation}
In the default setting the Hathor package is based on the LHAPDF
library~\cite{Whalley:2005nh} to access the PDFs.
The Hathor package has been tested with the most recent version 
lhapdf-5.8.3 
which can be obtained from {\tt http://projects.hepforge.org/lhapdf}. 
Please follow the instructions in the LHAPDF package to install the LHAPDF library.  
To build and use Hathor first unpack the package using 
{\tt tar xvfz Hathor.tar.gz} 
at the location where one wants to install the package. 
This will create a directory Hathor-1.0. Please
create a symbolic link {\tt lhapdf} inside this directory to the location of ones LHAPDF installation. 
The contents of {\tt lhapdf} should
contain the LHAPDF installation with the directories: {\tt bin include lib lib64 share}.
Then use make to build the Hathor package. 
Make will build the Hathor (static) library \verb+libHathor.a+ 
which can be used in other applications. 
In addition an example program \verb+main.exe+ is built. 
For details concerning the example program we refer to Section~\ref{sec:usage}. 
Note that the compilation is done using the GNU compilers gcc, gfortran and g++. 
The package is known to work also with the Intel compiler. 
In fact we recommend the usage of the Intel compiler since this results
in a much better performance. 
However, given that it is not everywhere available Hathor uses the GNU
compiler by default.~\footnote{For the Intel compiler the user has to adapt the {\tt makefile}.}

To run the example one has to tell the system where the dynamic
libraries for LHAPDF can be found. This is conveniently done using the
environment variable \verb#LD_LIBRARY_PATH#.  Using the C-shell the statement
would be:\\
\verb+setenv LD_LIBRARY_PATH <path_to_lhapdf_installation>/lib/+\\
In addition, one probably needs to specify the location where the 
grid files for the LHAPDF library are stored. Again using the C-shell, 
the statement would be of the form:\\
\verb+setenv LHAPATH <path_to_lhapdf_installation>/lhapdf/share/lhapdf/PDFsets+\\
For a detailed description concerning the paths required by the LHAPDF library, we refer to the LHAPDF manual. 
As concluding remarks with respect to the PDF libraries we would like to point
out that Hathor does not try to handle  errors from the LHAPDF library. 
This is not possible since LHAPDF does not provide a detailed error handling. 
Also note that since LHAPDF is not thread safe the same is true for Hathor.

\bigskip

Hathor is also equipped with a graphical user interface (GUI) written in Java. 
The GUI makes use of dynamic libraries to access the Hathor library within Java.~\footnote{%
The same technology can be used to access the Hathor library from Mathematica or Maple. The authors may provide the respective
interfaces in a future update on demand.}
The dynamic library is built with the command {\tt  CreateJavaGui.csh} which is included in the Hathor package.
The command {\tt  CreateJavaGui.csh} creates the dynamic library {\tt libHathor4Java.so} 
and writes the executable file {\tt xhathor} which is used to start the GUI.  
Please note that {\tt xhathor} sets up various paths:  
i.e. the environment variable {\tt LHAPATH} is set to {\tt ./lhapdf/share/lhapdf} 
if it has not been set yet. 
In addition the dynamic libraries {\tt libLHAPDF.so} and {\tt libgfortran.so}
on which the Hathor library relies are preloaded. 
This step is platform dependent and not alway easy to achieve in a universal way. 
If the graphical interface does not start with {\tt xhathor} the correct
setting of the paths is a likely source of potential errors. 
In that case we recommend to set the necessary paths directly in {\tt xhathor} 
or to consult the authors for support.

%
%
\section{Description}
\label{sec:description}
The entire cross section is calculated inside the class Hathor. 
This is done in order to avoid any possible problem with names used in already existing codes. 
Inside this class, a two dimensional numerical integration is performed in
which the PDFs are convoluted with the hard scattering cross section. 
As a numerical integration procedure we use the Vegas algorithm~\cite{Lepage78}.~\footnote{%
Hathor uses Vegas code which is a C port of the original fortran version \cite{Lepage78}.}
Since Vegas is a Monte Carlo integration we need to provide random numbers. 
Those are obtained by using the Ranlux algorithm~\cite{Luscher:1993dy} and 
we use the implementation available from Ref.~\cite{ranlux}.

The Hathor class takes as constructor argument a reference to the PDF 
which should be used in the current cross section calculation. 
In the following we list the publically available function calls and option choices together with a short description.
\begin{itemize}
\item  \verb+Hathor(Pdf & pdf)+\\
  Constructor to build one instance of the Hathor class. The argument is an instance of the PDF. 
  In case that LHAPDF is used the corresponding definition would be:\\
  {\tt Lhapdf pdf("MSTW2008nnlo68cl");}\\
  to use the {\tt MSTW2008nnlo68cl} set.\\
  At first sight it might appear strange that we use an additional
  ``wrapper class'' as interface to LHAPDF. The idea behind this is
  to give the user the possibility to become independent from LHAPDF.
  This is achieved by inherting the class {\tt Lhapdf}  from the
  base class {\tt Pdf}. By inhering its own class from the base class
  the user can thus easily implement its own wrapper to whatever PDF
  set he wants to use.
  As an example the {\tt class MSTW} has been supplied, 
  which gives direct access to the MSTW set~\cite{Martin:2009iq}. 
  (Note that in the MSTW case the $\alpha_s$ value is set to 1 since the library does not
  provide a function to evaluate it. The user has thus provide its own
  $\alpha_s$.)
  We have observed that in some cases the original code provided with 
  the PDF sets is faster than what is provided by LHAPDF. 
  Since the evaluation of the PDFs represents a significant part of the
  calculation the usage of the original PDFs may speed up the entire
  calculation significantly.
\item  \verb+void printOptions()+\\
  Use this function to print the options currently selected via the
  routine {\tt setScheme()};
\item  \verb+void setScheme(unsigned int newscheme)+\\
  Sets the specific scheme in particular perturbative order and
  renormalization schemes. Possible options are:
  \begin{itemize}
  \item 
    \verb+Hathor::LO+ to switch on the leading-order contribution.
  \item    
    \verb+Hathor::NLO+ to switch on the individual NLO contribution.
  \item    
    \verb+Hathor::NNLO+ to switch on the individual NNLO contribution (see Section~\ref{sec:methods}).
  \end{itemize}
  Please also note that for the computation of the total cross section 
  up to e.g. NNLO accuracy, it is necessary to combine the options as in\\
    \verb+setScheme(Hathor::LO | Hathor::NLO | Hathor::NNLO);+\\
  Other possible options are:
  \begin{itemize}
  \item    
    \verb+Hathor::MS_MASS+ to use the mass renormalized in the \msbar\ scheme (see Section~\ref{sec:methods}).
  \item    
    \verb+Hathor::LOG_ONLY+ to keep only the logarithmically enhanced 
    terms (see Section~\ref{sec:methods}).
    Please note that this option requires also 
    \verb+Hathor::NNLO+ to be set.
  \item    
    \verb+Hathor::PDF_SCAN+ to switch on the evaluation of the PDF
    uncertainties. That is to say, the error PDFs are also integrated along with
    the central value. To save computing time one may set the accuracy
    with {\tt XS.setPrecision(Hathor::LOW)} to {\tt LOW} in this case.
    Care has to be taken, though, by the user to ensure that the default PDF uncertainty
    estimate as implemented in Hathor (asymmetric error) complies with the conventions of the
    respective PDF set, as e.g. some PDF sets provide only a symmetric error. 
    In this case, the additional option \verb+Hathor::PDF_SYM_ERR+ needs to be
    set. See also the discussion in Section~\ref{sec:methods}.
  \item    
    \verb+Hathor::PDF_SYM_ERR+ invokes symmetric PDF error, if foreseen by the
    convention of the PDF set.
  \end{itemize}
  Please note, that these options can again be combined, e.g. as in\\
    \verb+setScheme(Hathor::LO | Hathor::NLO | Hathor::MS_MASS);+
\item \verb+void setColliderType( COLLIDERTYPE type)+\\
  Sets the hadronic initial state. Use \verb+Hathor::PP+ to select
  proton--proton collisions and \verb+Hathor::PPBAR+ for
  proton--anti-proton collisions.
  The collider energies are set to the default values: 7 TeV in case
  of proton--proton collisions and 1960 GeV in case of
  proton--anti-proton
  collisions. If this is inappropriate the values can be changed using 
  the command \verb+void setSqrtShad(double ecms)+, 
  where the center of mass energy is provided in GeV.
\item \verb+void setSqrtShad(double ecms)+\\
  Sets the center-of-mass energy in GeV.
\item \verb+void setNf(int nf)+ \\
  Sets the number of massless flavors to {\tt nf}. 
  For top-quark physics the default setting is $\nf=5$ and should not be changed. 
  This function may be used when the cross section for a hypothetical heavy
  quark of a fourth family is computed, 
  as Hathor includes the full $\nf$ dependence of the hard scattering cross section, 
  i.e. it features the formulae for general $\nf$.
  However, please note that the PDFs usually provide $\alpha_s$ in the $\nf=5$ flavor scheme. 
  So the user should be careful with this option (and the interpretation of the results).
\item \verb+void setCqq(double tmp)+ \\
  Can be used to set the constant
  defined in Eq.~(\ref{eq:fqq20}) to a specific value (see Section~\ref{sec:methods}). 
  The default is $C_{qq}=0$.
\item \verb+void setCgg(double tmp)+ \\ 
  Can be used to set the constant defined in Eq.~(\ref{eq:fgg20}) to a specific value (see Section~\ref{sec:methods}). 
  The default is $C_{gg}=0$
\item \verb+void setPrecision(int n)+ \\
  Can be used to define the accuracy of the numerical integration performed by the Hathor package. 
  It sets the number of function evaluations used in the Monte Carlo integration. 
  In principle, the user can provide any reasonable integer value.

  Pre-defined values are: 
  {\tt Hathor::LOW}, {\tt Hathor::MEDIUM}, {\tt Hathor::HIGH}.
  We recommend {\tt LOW} for the PDF scan and {\tt MEDIUM} for 
  the central value. 
  This should be sufficient for most applications. 
  Please note that {\tt Hathor::LOW} should give already an accuracy at the percent level.
  For detailed comparisons of theoretical results {\tt HIGH} may be used. 
\item \verb+double getAlphas(double mur)+\\
  Returns the QCD coupling constant at the renormalization scale
  \verb+mur+ as provided by the (central) PDF.  
\item \verb+double getXsection(double m, double mur, double muf)+\\
  This function starts the cross section calculation for a given
  top-quark mass and the factorization/renormalization scales provided
  as arguments. Unless a specific scheme is set through
  {\tt setScheme} the default setting is used:\\
  \verb+Hathor::LO | Hathor::NLO | Hathor::NNLO+\\
  The cross section for the  central PDF is returned. 
  More information can be obtained through {\tt getResult}.
\item \verb+void getResult(int pdfset, double & integral, double & err)+ \\
  This function is used to obtain additional information after the
  cross section has been calculated for a specific setting
  of the renormalization/factorization scale using {\tt getXsection}.
  The integer value {\tt pdfset} specifies the respective PDF: 0 for
  the central value, and 1 to {\tt getPdfNumber()} for the respective
  error PDF. 
  The result for the central value and the numerical error from
  the numerical integration are returned through {\tt integral} and {\tt err}.
  Note that {\tt err} should always be negligible.
  If this is not the case the precision of the numerical
  integration should be increased through {\tt setPrecision}.
\item \verb+void getPdfErr(double & up, double \& down)+\\
  This function returns the PDF uncertainty, if the option 
  \verb+Hathor::PDF_SCAN+ has been used.
  By default, Hathor assumes an asymmetric PDF error convention. 
  In case of a symmetric one, the option \verb+PDF_SCAN+ 
  has to be combined with the option \verb+PDF_SYM_ERR+ 
  (see Section~\ref{sec:methods} and the discussion above).
\item \verb+int getPdfNumber()+ \\
  Returns the number of error PDFs currently in use. 
  If 0 is returned the option \verb+PDF_SCAN+ is not switched on 
  or the PDF set does not support error PDFs.
  E.g., in case of the PDF set {\tt mstw2008nnlo.68cl} the return value would be 40.
\item \verb+void sethc2(double)+ \\
  Can be used to change the value for $(hc)^2$ 
  which is used by Hathor to convert the cross sections from GeV$^{-2}$ to pico barn. 
  The default used by Hathor is:\\
  {\tt 0.389379323e+9}.
\end{itemize}

%
%
\section{Usage and examples}
\label{sec:usage}
A concrete instance of the Hathor class is built using:
\begin{verbatim}
Lhapdf pdf("MSTW2008nnlo68cl");
Hathor XS(pdf);
\end{verbatim}
The evaluation of the cross section (using the default setting) 
is then done using
\begin{verbatim}
XS.getXsection(171.,171.*2,171./2);
\end{verbatim}
where the mass has been set to 171 GeV and $\mur=2\times171$ GeV and 
$\muf=171/2$ GeV. The result of the evaluation is obtained through
\begin{verbatim}
XS.getResult(0,val,err,chi2a);
\end{verbatim}
Note that {\tt XS.getXsection(171.,171.*2,171./2);} triggers the
numerical integration of the cross section. It has to be called first
before {\tt XS.getResult(0,val,err,chi2a);} can be used.
In a typical application we may want to use a lower statistic in the Monte
Carlo integration for the evaluation of the PDF uncertainty. 
This is achieved by the following code:
\begin{verbatim}
unsigned int scheme = Hathor::LO | Hathor::NLO | Hathor::NNLO;
double mt = 171., muf=171.,mur=171.;
double val,err,chi2a,up,down;

Lhapdf pdf("MSTW2008nnlo68cl");
Hathor XS(pdf);

XS.setPrecision(Hathor::MEDIUM);
XS.getXsection(mt,mur,muf);
XS.getResult(0,val,err,chi2a);

XS.setScheme(scheme | Hathor::PDF_SCAN);
XS.setPrecision(Hathor::LOW);
XS.getXsection(mt,mur,muf);
XS.getPdfErr(up,down);
\end{verbatim}
The central value is calculated with the precision set to {\tt MEDIUM}. 
The estimate of the PDF uncertainty is then obtained with a lower accuracy.

An example of the usage of Hathor illustrates the calculation with
running a mass. 
It reproduces the central curve (NNLO) of the right plot in Figure~5 of Ref.~\cite{Langenfeld:2009wd}. 
\begin{verbatim}
double val,err,chi;

Lhapdf lhapdf("MSTW2008nnlo68cl");
Hathor XS(lhapdf);

XS.setColliderType(Hathor::PPBAR);
XS.setScheme(Hathor::LO | Hathor::NLO | Hathor::NNLO | Hathor::MS_MASS );
XS.setPrecision(Hathor::LOW);

for(double mt = 140; mt < 181.; mt++ ){
  XS.getXsection(mt,mt,mt);
  XS.getResult(0,val,err,chi);
  cout << mt << " " << XS.getAlphas(mt) <<" "<< val << " " << err << endl;
}
\end{verbatim}

The typical runtimes for these examples range between seconds and a few
minutes and also depend on the chosen compilers.
E.g. on an Intel 3.00GHz QuadCore PC we have obtained 
with the options {\tt NNLO}, {\tt PDF\_SCAN} (PDF set {\tt MSTW2008nnlo68cl}) 
and {\tt XS.setPrecision(Hathor::LOW)} the result for the cross section after 
64 seconds using the gfortran compiler, and, 41 seconds respectively, using Intel's ifort compiler.

\bigskip
\bigskip

\begin{figure}[ht!]
\centering
\includegraphics[width=8.0cm,angle=270]{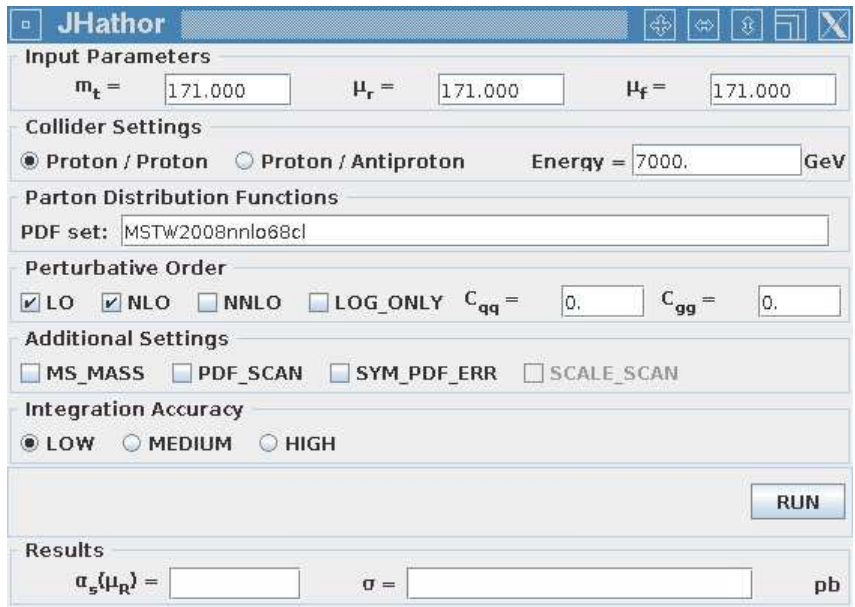}
\caption{\small
  \label{fig:JHathor}
  Screen shot of the Java graphical user interface for Hathor. 
}
\end{figure}

The Java GUI is invoked by the command {\tt xhathor} (see the discussion in
Sec.~\ref{sec:installation} for the installation). 
A screenshot is displayed in Fig.~\ref{fig:JHathor} and 
the input is self-explanatory with the help of  Sec.~\ref{sec:description} 
for the description of all options.
%

%
%
\section{Conclusions} 
\label{sec:conclusions}

Top-quark production at hadron colliders is at the edge of becoming a precision measurement 
requiring accurate theory predictions. 
Hathor is a fast and flexible program for the computation of the total cross section of
hadronic heavy-quark pair-production.
It takes into account the latest theoretical developments through a variety of options, 
it allows for separate variations of all scales and can be used with a large number of PDFs
through the LHAPDF interface.
As a novelty, Hathor offers predictions in
different renormalization schemes for the heavy-quark mass (pole and $\msbar$) 
and it can also be applied to a hypothetical fourth quark family assuming standard QCD couplings.
Thus, with these functionalities Hathor can serve as a reference for future cross section calculations. 

Hathor typically runs in a few seconds up to a few minutes (depending on the chosen options, e.g. extensive PDF scans)
on standard desktop or notebook PCs.
The Hathor package can either be used as a stand alone program or, as a small library, 
it can be easily integrated into existing code, e.g. for experimental analyses.

\smallskip

Hathor is publicly available for download from~\cite{hathor:2010} or from the authors upon request.

\subsection*{Acknowledgments}
We acknowledge useful discussions with S.~Alekhin and U.~Husemann.
This work is supported in part by the Deutsche Forschungsgemeinschaft in Sonderforschungs\-be\-reich/Transregio~9 
and the Helmholtz Gemeinschaft under contract VH-NG-105 
and contract VH-HA-101 (Alliance {\it ``Physics at the Terascale''}).

\appendix
%
%
\renewcommand{\theequation}{A.\arabic{equation}}
\setcounter{equation}{0}
\section{Total cross section with a running mass}
\label{sec:msbar}

The starting point is the relation between the on-shell mass and the \msbar\ mass:
\begin{equation}
\label{eq:pole-to-msbar}
  \mt = \mmu \* \left(1 + a_s \d1 + a_s^2 \d2\right)
  \, ,
\end{equation}
with $a_s = \as/\pi$.
If the decoupling $\as^{\nf=6}\to \as^{\nf=5}$ is performed at $\mmu$ the coefficients are given by
\begin{eqnarray} 
  \label{eq:d1def}
  \d1 &=& {4\over 3} + \ell 
  \, ,\\
  \label{eq:d2def}
  \d2 &=& {307\over 32} + 2 \* \z2 + {2\over 3} \* \z2 \* \ln 2 
  - {1\over 6}\*\z3 + {509\over 72}\*\ell + {47\over 24}\*\ell^2 
  \\
  && 
  - \left( {71\over 144} + {1\over 3}\*\z2 + {13\over 36}\*\ell 
    + {1\over 12}\*\ell^2 \right)\*\nf 
  + {4\over 3}\sum_l\Delta(m_l/\mt)
  \, ,
  \nn
\end{eqnarray}
which are known from Refs.~\cite{Gray:1990yh,Chetyrkin:1999qi,Melnikov:2000qh}
and $\ell = \ln(\mur^2 / \mmu^2)$. 
Note that the coefficients $d_i$ depend in general on the renormalization scale. 
Using $\mm$ instead of $\mmu$ the formulae simplify significantly. 
The full renormalization scale dependence can be restored at the end using renormalization group arguments. 
$\Delta(m_i/\mt)$ accounts for all massive quarks $m_i$ lighter than the top-quark. 
For all light quarks we set $m_l=0$ so the sum in Eq.~(\ref{eq:d2def}) vanishes. 

To convert the cross section to the \msbar\ mass scheme we start
from the hadronic cross section expanded in $\as$:
\begin{equation}
  \sigma_{h_1 h_2 \to {t\bar t X}}(S,\mt)
  = a_s^2 \sigma^{(0)}(S,\mt) 
  + a_s^3 \sigma^{(1)}(S,\mt) 
  + a_s^4 \sigma^{(2)}(S,\mt) 
  + O(a_s^5) 
  \, .
\end{equation}
Expressing $\mt$ in terms of $\mm$ and expanding in $\as$ we obtain
\begin{eqnarray}
  \label{eq:master}
  \sigma_{h_1 h_2 \to {t\bar t X}} &=& a_s^2 \* \sig0(S,\mm) + a_s^3 \* \sig1(S,\mm) + a_s^4\* \sig2(S,\mm)
  \\
  & & 
  + a_s^3 \*  \d1 \* \mm \* \Dsigma0 
  + a_s^4\*\left\{
    \mm \* \d2 \* \Dsigma0
  \right.
  \nn\\
  & &
  \left.
  + \mm \* \d1 \* \Dsigma1
  + {1\over 2}\*\mm^2\*\d1^2\* \DDsigma0
  \right\}
  \, .
  \nn
\end{eqnarray}
The derivatives of the LO cross sections with respect to
the mass can be written in the following form:
\begin{eqnarray}
  \label{eq:LOdiff1}
  \Dsigma0
  =
  {2\over \mm}\int_{4\mm^2}^S ds\,
  \left(s \, {d\over ds} \Lij(s,S,\muf)\right)
  \hat\sigma_{ij}^{(0)}(s,\mm)
  \, ,
\end{eqnarray}
and
\begin{eqnarray}
  \label{eq:LOdiff2}
 \left.{d^2\sigma^{(0)}\over d\mts}\right|_{\mt=\mm}
  &=& 
   -{2\over \mm^2}\int_{4\mm^2}^S ds  
   \left(s \, {d\over ds} \Lij(s,S,\muf)\right)
   \hat\sigma_{ij}^{(0)}(s,\mm)
   \\
  & &
  +\, {2\over \mm}\int_{4\mm^2}^S ds  
  \left(s \, {d\over ds} \Lij(s,S,\muf)\right)
  \left. {d\hat\sigma_{ij}^{(0)}\over d\mt}\right|_{\mt=\mm}
  \, ,
  \nn
\end{eqnarray}
where a summation over the contributing parton channels is understood.
Note that in Eqs.~(\ref{eq:master}),~(\ref{eq:LOdiff1}) and (\ref{eq:LOdiff2})
the renormalization scale is set to $\mur=\mm$.
The required derivatives of the LO partonic cross sections
with respect to the mass are easily obtained from Eqs.~(\ref{eq:LOqq}),~(\ref{eq:LOgg}):
\begin{eqnarray}
  \mt{d\sigma_{qq}\over d\mt} &=&
  - {1\over \mts} {1\over 9} \pi\as^2 {\rho^3\over \beta}
\, ,
\\
  \mt{d\sigma_{gg}\over d\mt} &=&
  {1\over 192}\*\pi\*\as^2\*{1\over \mts}\*{\rho\over \beta}
  \left( \beta\*(36 - 40\beta^2 + 4\*\beta^4)
    \*\ln\left({1+\beta\over 1-\beta}\right)
    -7 -116\*\beta^2 + 91\*\beta^4
  \right)
  \, .
  \quad
\end{eqnarray}
For the first derivative of $\sigma^{(1)}$ we obtain a similar result:
\begin{eqnarray}
  \left.{d\sigma^{(1)}\over d\mt}\right|_{\mt=\mm} 
  &=& - \int_{4\mm^2}^S ds \Lij( s,S, \muf)
  {1\over \mm}\tilde\sigma_{ij}^{(1)}(s,\mm)
  \\
  & &+\, {2\over \mm}\int_{4\mm^2}^S ds
  \left(s \, {d\over ds} 
  \Lij(s,S,\muf)\right)\sigma^{(i)}\left(s,\mm,{\muf\over\mm},1\right)
\, ,
\nn
\end{eqnarray}
with 
\begin{equation}
  \tilde\sigma_{ij}^{(1)}(s,\mm) =
  \muf {\partial \over \partial \muf}
  \sigma_{ij}^{(1)}\left(s,\mm,{\muf\over \mm},1\right)
  \, .
\end{equation}
Using Eq. (\ref{eq:DefinitionScalingFunctions}) 
the contribution $\tilde\sigma_{ij}^{(1)}(s,\mm)$ can be written as
\begin{equation}
  a_s^2\tilde \sigma^{(1)}_{ij} = 8 {\as^2(\mm)\over\mm^2} \fijoo
  \, .
\end{equation}
Since the luminosities are only known numerically the derivatives are evaluated using
\begin{equation}
  {d\over ds} \Lij(s,S,\muf) = 
  {1\over 2\delta}\left(\Lij(s+\delta,S,\muf) -
  \Lij(s-\delta,S,\muf)\right)
 + O(\delta^2)
 \, .
\end{equation}
The results presented so far are only valid for $\mur = \mm$. 
Using
\begin{equation}
  a_s(\mm) = a_s(\mur)\*\left( 1 
    +  4\pi^2 a_s(\mur)\*L_{\bar R} \* \b0 
    + (4\pi^2)^2 a_s(\mur)^2\* ( \b1\*L_{\bar R} + \b0^2 \*L_{\bar R}^2)\right)
  \, ,
\end{equation}
with $L_{\bar R} = \ln(\mur^2 / \mm^2)$ 
it is easy to restore the complete renormalization scale dependence.

%
%
\renewcommand{\theequation}{B.\arabic{equation}}
\setcounter{equation}{0}
\section{Scaling functions}
\label{sec:fits}

Here we present the expressions for the scaling functions as implemented in
the program Hathor. 
At Born level,
\begin{eqnarray}
\label{eq:fqq0}
\fqqn &=& 
{\pi\, \beta\, \rho \over 27} [2 + \rho]
\, ,
\\
\label{eq:fgq0}
\fgqn &=& 
0
\, ,
\\
\label{eq:fgg0}
\fggn &=& 
{\pi\, \beta\, \rho \over 192} \biggl[
{1 \over \beta} (\rho^2 + 16 \rho + 16)  \ln\biggl(\frac{1+\beta}{1-\beta}\biggr)
-28 - 31 \rho
\biggr]
\, ,
\end{eqnarray}
where $\beta=\sqrt{1-\rho}$ and $\rho = 4 \mts /s$.
At NLO the functions $\fijon$ have been described through 
precise parametrizations with per mil accuracy and the following ansatz~\cite{Langenfeld:2009wd}:
\begin{eqnarray}
\label{eq:fqq10}
\fqqon &=& 
   \frac{\rho \* \beta}{36\pi} \biggl[
     \frac{32}{3} \ln^2 \beta
   + \biggl(32 \ln 2 - \frac{82}{3}\biggr)\, \ln \beta
   - \frac{1}{12}\*\frac{\pi^2}{\beta}
   \biggr]
   + \beta  \rho a_0^{qq} + h(\beta,a_1,\ldots,a_{17}) 
\\
& &
   + \frac{1}{8\pi^2}(\nf-4) \fqqn \biggl[
     \frac{4}{3}  \ln 2 - \frac{2}{3} \ln \rho
   - \frac{10}{9}
   \biggr]
\ ,
\nonumber\\
\label{eq:fgq10}
\fgqon &=& 
   \frac{1}{16\pi}\beta^3\biggl[
   \frac{5}{9} \ln \beta
   + \frac{5}{6}\ln 2 - \frac{73}{108} 
    \biggr] 
   + h_{gq}^{(a)}(\beta,a_1,\ldots,a_{15})
\ ,
\\
\label{eq:fgg10}
\fggon &=& 
\frac{7\beta}{768\pi} \biggl[
     24 \ln^2 \beta
   +\biggl(72 \ln 2 - \frac{366}{7} \biggr)\ln \beta
   + \frac{11}{84}\*\frac{\pi^2}{\beta}
   \biggr] 
   + \beta  a_0^{gg} + h(\beta,a_1,\ldots,a_{17})
\\ & &
   + (\nf-4)\frac{\rho^2}{1024\pi} \biggl[
      \ln\biggl(\frac{1+\beta}{1-\beta}\biggr)
   - 2 \beta 
   \biggr]
\, ,
\nonumber
\end{eqnarray}
where $\nf$ denotes the number of light quarks and the complete $\nf$-dependence has been kept manifest.
The constants $a_0^{ij}$ read $a_0^{qq}= 0.03294734$ and $a_0^{gg} = 0.01875287$ 
and the fit functions $h(\beta,a_1,\ldots,a_{17})$ and $h_{gq}^{(a)}(\beta,a_1,\ldots,a_{15})$ 
are given in Eqs.~(\ref{eq:tophgg}), (\ref{eq:tophgqa}) together 
with a list of all parameters in Tabs.~\ref{tab:qqfit}--\ref{tab:ggfit}.
The exact expressions for scale dependent functions $\fijoo$ have already compact analytical form 
containing at most dilogarithms. 
They read~\cite{Nason:1987xz}
\begin{eqnarray}
\label{eq:fqq11}
\fqqoo &=& \frac{1}{8\*\pi^2}\*\left[
                        \frac{16\*\pi}{81} \* \rho \* \ha
                        +\frac{1}{9} \fqqn(\rho) \* \left(
                        127 - 6\*\nf+48 \*\hb \right)
                        \right]
                        \, , 
                        \\
\label{eq:fgq11}
\fgqoo &=& \frac{1}{8\*\pi^2}\*\frac{\pi}{192}\*\left[
                  \frac{4}{9}\*\rho\*\left(14\*\rho^2+27\*\rho-136\right)\*\ha\right. 
                \\
                && \hspace*{20mm}\left.
                  -\frac{32}{3}\*\rho\*\left(2-\rho\right)\*h_1(\beta)
                  -\frac{8}{135}\*\beta\*\left(1319\*\rho^2 - 3468\*\rho + 724\right)
                        \right]
                        \, ,
                        \nn\\
\label{eq:fgg11}
\fggoo &=& \frac{1}{8\*\pi^2}\*\left[
                  \frac{\pi}{192}\*\left\{
                  2\*\rho\left(59\*\rho^2+198\*\rho -288\right)\* \ha
                \right.\right. \\
                  && \hspace*{20mm}\left.\left.
                  +12\*\rho\*\left(\rho^2+16\*\rho+16\right)\*h_2(\beta) 
                  -6\*\rho\left(\rho^2-16\*\rho+32\right)\*h_1(\beta)
                \right.\right.
              \nn\\
                  && \hspace*{20mm}\left.\left.
                  -\frac{4}{15}\*\beta\*\left(7449\*\rho^2-3328\*\rho+724\right)
                  \right\}
                  +12\*\fggn(\rho)\hb
                  \right] 
                  \, , 
                  \nn
\end{eqnarray}
with the auxiliary functions containing the standard dilogarithm $\Liz(x)= -\int_0^x\frac{\mathrm{d}t}{t}\ln(1-t)$,
\begin{eqnarray}
h_1(\beta) &=& \ln^2\left(\argp\right) -  \ln^2\left(\argm\right) 
              + 2\*\Liz\left(\argp\right) - 2\*\Liz\left(\argm\right) \, , \\
h_2(\beta) &=&\Liz\left(\argpp\right) - \Liz\left(\argmm\right) \, .
\end{eqnarray}

At NNLO the functions $\fijto$ and $\fijtt$ are known exactly~\cite{Langenfeld:2009wd}.
The fits to the scaling functions generally have per mil accuracy with  
exceptions in regions close to zero, where an accuracy better than one percent is kept.
\begin{eqnarray}
\label{eq:f21qq}
\fqqto &=& 
  \frac{1}{(16\pi^2)^2} \fqqn \biggl[
  - \frac{8192}{9}\* \ln^3 \beta 
  + \biggl( \frac{12928}{3} - \frac{32768}{9}\* \ln 2 \biggr)\* \ln^2 \beta
\\
&& 
  + \biggl( - 840.51065 + 70.183854 \frac{1}{\beta} \biggr)\* \ln \beta
  - 82.246703\frac{1}{\beta} + 467.90402 
  \biggr] 
\nonumber\\
&&
  + \frac{\nf}{(16\pi^2)^2} \fqqn \biggl[
  - \frac{256}{3}\* \ln^2 \beta
  + \biggl( \frac{2608}{9} - \frac{2816}{9}\* \ln 2 \biggr) \* \ln \beta
  + 6.5797363\frac{1}{\beta} - 64.614276\biggr] 
\nonumber\\
&& 
  + h(\beta,b_i + \nf \*c_i)
  - \frac{4 \nf^2}{(16\pi^2)^2} \fqqn \left[
    \frac{4}{3} \ln 2 - \frac{2}{3} \ln \rho - \frac{10}{9}\right]
\, ,
\nonumber\\
\label{eq:f22qq}
\fqqtt &=& 
  \frac{1}{(16\pi^2)^2} \fqqn \biggl[
  \frac{2048}{9} \* \ln^2 \beta 
  + \biggl( - \frac{7840}{9} + \frac{4096}{9}\* \ln 2 \biggr) \* \ln \beta  
  + 270.89724
  \biggr]
\\
&& 
  + \frac{\nf}{(16\pi^2)^2} \fqqn \biggl[
    \frac{320}{9}\* \ln \beta 
  - \frac{596}{9} 
  + \frac{320}{9}\* \ln 2
  \biggr] 
  + h(\beta,b_i + \nf \*c_i)
  + \frac{4 \nf^2}{3 (16\pi^2)^2} \fqqn 
\, ,
\nonumber\\
\label{eq:f21gq}
\fgqto &=&
  - \frac{\pi}{(16\pi^2)^2} \beta^3 \biggl[ 
    \frac{770}{27} \ln^2 \beta
  + \biggl( - \frac{6805}{81} + \frac{6160}{81} \ln 2 \biggr) \ln \beta 
  + 0.13707784\frac{1}{\beta} 
\\
&& 
  + 0.22068868 
  \biggr]
- \frac{\pi \nf}{81(16\pi^2)^2} \beta^3 \biggl[
  46 \ln \beta - \frac{163}{3} + 76 \ln 2 
  \biggr]
  + h_{gq}^{(b)}(\beta,b_i + \nf \*c_i)
\nonumber\\
\label{eq:f22gq}
\fgqtt &=&
  \frac{\pi}{(16\pi^2)^2} \beta^3 \biggl[
  \frac{385}{81} \ln \beta - \frac{1540}{243} + \frac{385}{81} \ln 2 
  \biggr] 
  + h_{gq}^{(b)}(\beta,b_i + \nf \*c_i)
\, ,
\\
\label{eq:f21gg}
\fggto &=& 
  \frac{1}{(16\pi^2)^2} \fggn \biggl[
  - 4608\* \ln^3 \beta 
  + \biggl(\frac{109920}{7}-18432 \* \ln 2 \biggr)\* \ln^2 \beta
\\
&&
  + \biggl( 69.647185 - 248.15005 \frac{1}{\beta} \biggr) \* \ln \beta
  + 56.867721 \frac{1}{\beta} + 17.010070
  \biggr] 
\nonumber\\ 
&&
  + \frac{\nf}{(16\pi^2)^2} \fggn \biggl[
  - 64 \* \ln^2 \beta
  + \biggl( \frac{4048}{21} - 192 \* \ln 2 \biggr)\* \ln \beta
  - 3.4465285 \frac{1}{\beta} - 37.602004\biggr]
\nonumber\\
&& 
  + h(\beta,b_i + \nf \*c_i)
  \, ,
\nonumber\\
\label{eq:f22gg}
\fggtt &=& 
  \frac{1}{(16\pi^2)^2} \fggn \biggl[
    1152 \* \ln^2 \beta 
  + ( - 2568 + 2304 \* \ln 2 )\* \ln \beta 
  - 79.74312140 
  \biggr]
\\
&&
  + \frac{\nf}{(16\pi^2)^2} \fggn \biggl[
    16\* \ln \beta 
  - 16 + 16\* \ln 2
  \biggr] 
  + h(\beta,b_i + \nf \*c_i)
\, ,
\nonumber
\end{eqnarray}
with the fit functions $h(\beta,a_1,\ldots,a_{17})$  and $h_{gq}^{(b)}(\beta,a_1,\ldots,a_{18})$ 
(see Tabs.~\ref{tab:qqfit}--\ref{tab:ggfit} for a list of all parameters),
\begin{eqnarray}
\label{eq:tophgg}
h(\beta,a_1,\ldots,a_{17})&=& 
    a_{{1}}{\beta}^{2}
  + a_{{2}}{\beta}^{3}
  + a_{{3}}{\beta}^{4}
  + a_{{4}}{\beta}^{5}
\\
&&
  + a_{{5}}{\beta}^{2} \ln \beta 
  + a_{{6}}{\beta}^{3} \ln \beta 
  + a_{{7}}{\beta}^{4} \ln \beta 
  + a_{{8}}{\beta}^{5} \ln \beta 
\nonumber\\
&&  
  + a_{{9}}{\beta}^{2}  \ln^{2} \beta
  + a_{{10}}{\beta}^{3} \ln^{2} \beta
  + a_{{11}}\beta\, \ln \rho 
  + a_{{12}}\beta\,  \ln^{2} \rho
  + a_{{13}}{\beta}^{2} \ln \rho
\nonumber\\
&&  
  + a_{{14}}{\beta}^{2}  \ln^{2} \rho
  + a_{{15}}{\beta}^{3} \ln \rho 
  + a_{{16}}{\beta}^{3}  \ln^{2} \rho
  + a_{{17}}{\beta}^{4} \ln \rho 
\, ,
\nonumber\\
\label{eq:tophgqa}
h_{gq}^{(a)}(\beta,a_1,\ldots,a_{15}) &=&
    a_{{1}}{\beta}^{4}
  + a_{{2}}{\beta}^{5}
  + a_{{3}}{\beta}^{6}
\\
&&
  + a_{{4}}{\beta}^{4} \ln \beta 
  + a_{{5}}{\beta}^{5} \ln \beta 
  + a_{{6}}{\beta}^{6} \ln \beta 
\nonumber\\
&&
  + a_{{7}}{\beta}^{2}\rho\, \ln \rho 
  + a_{{8}}{\beta}^{2}\rho\,  \ln^2 \rho
  + a_{{9}}{\beta}^{3}\rho\, \ln \rho
\nonumber\\
&&
  + a_{{10}}{\beta}^{3}\rho\, \ln^2 \rho
  + a_{{11}}{\beta}^{4}\rho\, \ln \rho 
\nonumber\\
&& 
  + a_{{12}}{\beta}^{4}\rho\, \ln^2 \rho
  + a_{{13}}{\beta}^{2}\rho\, \ln^3 \rho
  + a_{{14}}{\beta}^{2}\rho\, \ln^4 \rho
  + a_{{15}}{\beta}^{2}\rho\, \ln^5 \rho
\, ,
\nonumber\\
\label{eq:tophgqb}
h_{gq}^{(b)}(\beta,a_1,\ldots,a_{18})&=&
    a_{{1}}{\beta}^{3}
  + a_{{2}}{\beta}^{4}
  + a_{{3}}{\beta}^{5}
  + a_{{4}}{\beta}^{6}
  + a_{{5}}{\beta}^{7}
\\
&&
   + a_{{6}}{\beta}^{4} \ln \beta
   + a_{{7}}{\beta}^{5} \ln \beta
   + a_{{8}}{\beta}^{6} \ln \beta
   + a_{{9}}{\beta}^{7} \ln \beta
\nonumber\\
&&
   + a_{{10}}{\beta}^{3} \ln \rho
   + a_{{11}}{\beta}^{3} \ln^2 \rho
   + a_{{12}}{\beta}^{4} \ln \rho
   + a_{{13}}{\beta}^{4} \ln^2 \rho
\nonumber\\
&& 
   + a_{{14}}{\beta}^{5} \ln \rho
   + a_{{15}}{\beta}^{5} \ln^2 \rho
   + a_{{16}}{\beta}^{6} \ln \rho
   + a_{{17}}{\beta}^{6} \ln^2 \rho
   + a_{{18}}{\beta}^{7} \ln \rho
\, .
\nonumber
\end{eqnarray}
\begin{table}[ht!]
\centering
\begin{tabular}{r|r|rr|rr}
  &\multicolumn{1}{c|}{$\fqqon$} &\multicolumn{2}{c|}{$\fqqto$} &\multicolumn{2}{c}{$\fqqtt$}\\[1mm]
  $i$&\multicolumn{1}{c|}{$a_i$} & \multicolumn{1}{c}{$b_i$} & \multicolumn{1}{c|}{$c_i$} &
   \multicolumn{1}{c}{$b_i$} & \multicolumn{1}{c}{$c_i$} \\
\hline
$  1 $ & $    0.07120603  $ & $   -0.15388765  $ & $    -0.07960658  $ & $    0.37947056 $ & $   -0.00224114 $  \\[1mm]
$  2 $ & $   -1.27169999  $ & $    4.85226571  $ & $     0.50111294  $ & $   -4.25138041 $ & $    0.02685576 $  \\[1mm]
$  3 $ & $    1.24099536  $ & $   -7.06602840  $ & $    -0.09496432  $ & $    2.91716094 $ & $   -0.01777126 $  \\[1mm]
$  4 $ & $   -0.04050443  $ & $    2.36935255  $ & $    -0.32590203  $ & $    0.94994470 $ & $   -0.00626121 $  \\[1mm]
$  5 $ & $    0.02053737  $ & $   -0.03634651  $ & $    -0.02229012  $ & $    0.10537529 $ & $   -0.00062062 $  \\[1mm]
$  6 $ & $   -0.31763337  $ & $    1.25860837  $ & $     0.23397666  $ & $   -1.69689874 $ & $    0.00980999 $  \\[1mm]
$  7 $ & $   -0.71439686  $ & $    2.75441901  $ & $     0.30223487  $ & $   -2.60977181 $ & $    0.01631175 $  \\[1mm]
$  8 $ & $    0.01170002  $ & $   -1.26571709  $ & $     0.13113818  $ & $   -0.27215567 $ & $    0.00182500 $  \\[1mm]
$  9 $ & $    0.00148918  $ & $   -0.00230536  $ & $    -0.00162603  $ & $    0.00787855 $ & $   -0.00004627 $  \\[1mm]
$ 10 $ & $   -0.14451497  $ & $    0.15633927  $ & $     0.08378465  $ & $   -0.47933827 $ & $    0.00286176 $  \\[1mm]
$ 11 $ & $   -0.13906364  $ & $    1.79535231  $ & $    -0.09147804  $ & $   -0.18217132 $ & $    0.00111459 $  \\[1mm]
$ 12 $ & $    0.01076756  $ & $    0.36960437  $ & $    -0.01581518  $ & $   -0.04067972 $ & $    0.00017425 $  \\[1mm]
$ 13 $ & $    0.49397845  $ & $   -5.45794874  $ & $     0.26834309  $ & $    0.54147194 $ & $   -0.00359593 $  \\[1mm]
$ 14 $ & $   -0.00567381  $ & $   -0.76651636  $ & $     0.03251642  $ & $    0.08404406 $ & $   -0.00035339 $  \\[1mm]
$ 15 $ & $   -0.53741901  $ & $    5.35350436  $ & $    -0.25679483  $ & $   -0.51918414 $ & $    0.00363300 $  \\[1mm]
$ 16 $ & $   -0.00509378  $ & $    0.39690927  $ & $    -0.01670122  $ & $   -0.04336452 $ & $    0.00017915 $  \\[1mm]
$ 17 $ & $    0.18250366  $ & $   -1.68935685  $ & $     0.07993054  $ & $    0.15957988 $ & $   -0.00115164 $  \\[1mm]
\end{tabular}
\small
\caption{\small
\label{tab:qqfit}
Coefficients for fits of the $q\bar{q}$ scaling functions.}
\end{table}
\begin{table}[ht!]
\centering
\begin{tabular}{r|r|rr|rr}
  &\multicolumn{1}{c|}{$\fgqon$} &\multicolumn{2}{c|}{$\fgqto$} &\multicolumn{2}{c}{$\fgqtt$}\\[1mm]
  $i$&\multicolumn{1}{c|}{$a_i$} & \multicolumn{1}{c}{$b_i$} & \multicolumn{1}{c|}{$c_i$} &
   \multicolumn{1}{c}{$b_i$} & \multicolumn{1}{c}{$c_i$} \\
\hline
$  1 $ & $   -0.26103970  $ & $   -0.00120532  $ & $     0.00003257  $ & $   -0.00022247 $ & $    0.00001789 $  \\[1mm]
$  2 $ & $    0.30192672  $ & $   -0.04906353  $ & $     0.00014276  $ & $    0.00050422 $ & $    0.00000071 $  \\[1mm]
$  3 $ & $   -0.01505487  $ & $   -0.20885725  $ & $    -0.00402017  $ & $   -0.02945504 $ & $   -0.00020581 $  \\[1mm]
$  4 $ & $   -0.00142150  $ & $  -13.73137224  $ & $     0.06329831  $ & $    0.34340412 $ & $    0.00108759 $  \\[1mm]
$  5 $ & $   -0.04660699  $ & $   14.01818840  $ & $    -0.05952825  $ & $   -0.31894917 $ & $   -0.00086284 $  \\[1mm]
$  6 $ & $   -0.15089038  $ & $   -0.00930488  $ & $     0.00002694  $ & $    0.00009213 $ & $    0.00000010 $  \\[1mm]
$  7 $ & $   -0.25397761  $ & $   -0.52223668  $ & $     0.00159804  $ & $    0.00690402 $ & $    0.00001638 $  \\[1mm]
$  8 $ & $   -0.00999129  $ & $   -4.68440515  $ & $     0.01522672  $ & $    0.07847233 $ & $    0.00022730 $  \\[1mm]
$  9 $ & $    0.39878717  $ & $   -7.61046166  $ & $     0.02869438  $ & $    0.16042051 $ & $    0.00045698 $  \\[1mm]
$ 10 $ & $   -0.02444172  $ & $    1.36687743  $ & $    -0.00875589  $ & $   -0.05186974 $ & $   -0.00025620 $  \\[1mm]
$ 11 $ & $   -0.14178346  $ & $    1.84698291  $ & $    -0.00800271  $ & $   -0.03861021 $ & $   -0.00016026 $  \\[1mm]
$ 12 $ & $    0.01867287  $ & $   -7.26265988  $ & $     0.04043479  $ & $    0.21650362 $ & $    0.00070713 $  \\[1mm]
$ 13 $ & $    0.00238656  $ & $   -4.89364026  $ & $     0.01965878  $ & $    0.10137656 $ & $    0.00034937 $  \\[1mm]
$ 14 $ & $   -0.00003399  $ & $   11.04566784  $ & $    -0.05262293  $ & $   -0.28056264 $ & $   -0.00072547 $  \\[1mm]
$ 15 $ & $   -0.00000089  $ & $    4.13660190  $ & $    -0.01457395  $ & $   -0.08090469 $ & $   -0.00025525 $  \\[1mm]
$ 16 $ & $    0.00000000  $ & $   -6.33477051  $ & $     0.02314616  $ & $    0.13077889 $ & $    0.00034015 $  \\[1mm]
$ 17 $ & $    0.00000000  $ & $   -1.08995440  $ & $     0.00291792  $ & $    0.01813862 $ & $    0.00006613 $  \\[1mm]
$ 18 $ & $    0.00000000  $ & $    1.19010561  $ & $    -0.00220115  $ & $   -0.01585757 $ & $   -0.00006562 $  \\[1mm]
\end{tabular}
\caption{\small
\label{tab:gqfit}
Coefficients for fits of the $gq$ scaling functions.}
\end{table}
\begin{table}[ht!]
\centering
\begin{tabular}{r|r|rr|rr}
  &\multicolumn{1}{c|}{$\fggon$} &\multicolumn{2}{c|}{$\fggto$} &\multicolumn{2}{c}{$\fggtt$}\\[1mm]
  $i$&\multicolumn{1}{c|}{$a_i$} & \multicolumn{1}{c}{$b_i$} & \multicolumn{1}{c|}{$c_i$} & \multicolumn{1}{c}{$b_i$} & \multicolumn{1}{c}{$c_i$} \\
\hline
$  1 $ & $    -8.92563222  $ & $   -4.18931464  $ & $     0.12306772  $ & $    0.01222783 $ & $   -0.00380386 $  \\[1mm]
$  2 $ & $   149.90572830  $ & $   82.35066406  $ & $    -2.75808806  $ & $   -0.77856184 $ & $    0.08757766 $  \\[1mm]
$  3 $ & $  -140.55601420  $ & $  -87.87311969  $ & $     3.19739272  $ & $    1.33955698 $ & $   -0.10742267 $  \\[1mm]
$  4 $ & $    -0.34115615  $ & $    9.80259328  $ & $    -0.56233045  $ & $   -0.59108409 $ & $    0.02382706 $  \\[1mm]
$  5 $ & $    -2.41049833  $ & $   -1.12268550  $ & $     0.03240048  $ & $    0.00248333 $ & $   -0.00099760 $  \\[1mm]
$  6 $ & $    54.73381889  $ & $   29.51830225  $ & $    -0.92541788  $ & $   -0.23827213 $ & $    0.02932941 $  \\[1mm]
$  7 $ & $    90.91548015  $ & $   48.36110694  $ & $    -1.57154712  $ & $   -0.38868910 $ & $    0.04906147 $  \\[1mm]
$  8 $ & $    -4.88401008  $ & $   -7.06261770  $ & $     0.35109760  $ & $    0.28342153 $ & $   -0.01373734 $  \\[1mm]
$  9 $ & $    -0.17466779  $ & $   -0.08025226  $ & $     0.00227936  $ & $    0.00010876 $ & $   -0.00006986 $  \\[1mm]
$ 10 $ & $    13.47033628  $ & $    7.01493779  $ & $    -0.21030153  $ & $   -0.03383862 $ & $    0.00658371 $  \\[1mm]
$ 11 $ & $    22.66482710  $ & $   15.00588140  $ & $    -0.63688407  $ & $   -0.29071016 $ & $    0.02089321 $  \\[1mm]
$ 12 $ & $     4.60726682  $ & $    3.84142441  $ & $    -0.12959776  $ & $   -0.11473654 $ & $    0.00495414 $  \\[1mm]
$ 13 $ & $   -67.62342328  $ & $  -47.02161789  $ & $     1.91690216  $ & $    0.98929369 $ & $   -0.06553459 $  \\[1mm]
$ 14 $ & $    -9.70391427  $ & $   -8.05583379  $ & $     0.26755747  $ & $    0.24899069 $ & $   -0.01046635 $  \\[1mm]
$ 15 $ & $    65.08050888  $ & $   47.02740535  $ & $    -1.86154423  $ & $   -1.06096321 $ & $    0.06559130 $  \\[1mm]
$ 16 $ & $     5.09663260  $ & $    4.21438052  $ & $    -0.13795865  $ & $   -0.13425338 $ & $    0.00551218 $  \\[1mm]
$ 17 $ & $   -20.12225341  $ & $  -14.99599732  $ & $     0.58155056  $ & $    0.35935660 $ & $   -0.02095059 $  \\[1mm]
\end{tabular}
\caption{\small
\label{tab:ggfit}
Coefficients for fits of the $gg$ scaling functions.}
\end{table}

\cleardoublepage
%
%

{\footnotesize


}

\end{document}